\documentclass[11pt,letterpaper]{article}
\usepackage[margin=1in]{geometry}
\usepackage[T1]{fontenc}
\usepackage{amsmath,amssymb,amsthm,microtype}
\usepackage[hidelinks]{hyperref}
\newtheorem{problem}{Problem}
\newtheorem{theorem}{Theorem}
\newtheorem{lemma}{Lemma}

\newtheorem{example}{Example}
\newcommand{\excontname}{}
\newtheorem*{excontinner}{\excontname}
\newenvironment{excont}[1]
  {\renewcommand{\excontname}{Example~\ref*{#1}}\begin{excontinner}[continued]}
  {\end{excontinner}}

\newcommand{\F}{\mathbb F}
\newcommand{\ip}[2]{\langle #1,#2\rangle}
\usepackage[dvipsnames]{xcolor}

\begin{document}

\title{Making Every Edit Count\\
{\large Efficient Binary Edit Correction at the Gilbert--Varshamov Bound}}
\author{
Ryan Gabrys\thanks{\normalsize\raggedright Qualcomm Institute, University of California San Diego, San Diego, CA, USA. Email \href{mailto:rgabrys@ucsd.edu}{\texttt{rgabrys@ucsd.edu}}.}
\and
Shubhransh Singhvi\thanks{\normalsize\raggedright Department of Computer Science, Technion---Israel Institute of Technology, Haifa, Israel and Fuzzy Dynamics, Bengaluru, India. Email \href{mailto:shubhranshsinghvi2001@gmail.com}{\texttt{shubhranshsinghvi2001@gmail.com}}.}
\and
Eitan Yaakobi\thanks{\normalsize\raggedright Henry and Marilyn Taub Faculty of Computer Science, Technion---Israel Institute of Technology, Haifa, Israel. Email \href{mailto:yaakobi@cs.technion.ac.il}{\texttt{yaakobi@cs.technion.ac.il}}.\\
\textit{ -- Author names are listed alphabetically.}}
}
\date{}
\hypersetup{
  pdftitle={Making Every Edit Count - Efficient Binary Edit Correction at the Gilbert--Varshamov Bound},
  pdfauthor={Ryan Gabrys, Shubhransh Singhvi, Eitan Yaakobi},
  pdfsubject={Binary codes correcting insertions, deletions, and substitutions},
  pdfkeywords={error-correcting codes, edit errors, deletion correction, Gilbert--Varshamov bound}
}
\pagenumbering{roman}
\maketitle
\thispagestyle{empty}

\begin{abstract}
\normalsize
We construct binary codes that attain the classical Gilbert--Varshamov redundancy
guarantee for a fixed number of edit errors, with deterministic polynomial-time
construction, encoding, and decoding. For every fixed integer $t\ge 1$ and all
sufficiently large block lengths $n$, our codes correct any combination of at most
$t$ insertions, deletions, and substitutions in total, with redundancy at most
$2t\log_2 n+O_t(1)$. All algorithms run in polynomial time in $n$ for fixed $t$, and
decoding recovers the original information message. This realizes the classical
counting guarantee algorithmically without enumerating the exponentially large
space of binary words.

\end{abstract}
\clearpage
\pagenumbering{arabic}

\section{Introduction}

A handful of missing bits can create uncertainty far beyond the values of the bits that were lost.
Deletions disturb the alignment between a transmitted message and the sequence that
reaches the receiver, so that even the surviving symbols may no longer have identifiable
positions. This loss of synchronization turns a seemingly modest error model into a
fundamental challenge for reliable communication. Since the early 1960s, researchers have
sought codes that overcome it with minimal redundancy and efficient algorithms. The
history of deletion correction is marked by a striking tension. A nearly optimal solution
for a single deletion emerged early, while extending that success to multiple deletions
has demanded decades of new ideas.

The contrast with classical error correction explains the difficulty. A substitution
changes a symbol's value but preserves its position. An erasure leaves a marked gap. In
either case, the receiver retains the correspondence between transmitted and received
coordinates, even though some values are wrong or unknown. Familiar algebraic codes
exploit this correspondence through checks involving specified positions. An insertion or
deletion disrupts this correspondence, forcing the decoder to account for uncertainty
about alignment when interpreting such checks. Different alignments can make different
original messages consistent with the same received sequence. The code must therefore
control uncertainty about both symbol values and symbol positions, and classical
techniques do not transfer automatically~\cite{HS2021}.

File synchronization brings this problem into everyday computing. Machines and cloud
services must reconcile different versions of a file without retransmitting content they
already share. Insertions and deletions shift that content, so matching portions may
occur at different offsets. The widely used rsync utility addresses this challenge
through hashing, combining rolling checksums with stronger hashes to identify matching
blocks and reuse data already present at the receiver~\cite{TM1996}. Improving such
methods remains an active systems problem. SkySync, presented at FAST 2026, builds on
rsync and related protocols to reduce the computational cost of cloud
synchronization~\cite{SkySync2026}. Reducing communication creates a direct role for
deletion-correcting codes. Ni, Tauz, Gabrys, and Dolecek incorporate codes that correct
multiple deletions into an interactive synchronization protocol, reducing the communication
needed to reconcile files when one is a subsequence of the other. Their results show how
advances in deletion correction can translate into more efficient
synchronization~\cite{NTGD2025}.

The same challenge underlies both DNA data storage and robust watermarking of
AI-generated text. In DNA data storage, digital information is encoded in nucleotide
sequences, and synthesis and sequencing can introduce insertions, deletions, and
substitutions~\cite{HMG2019}. A watermark intended to identify a text's source, in turn,
must remain detectable when words are deleted, inserted, or changed. The 2024
SynthID-Text study evaluated watermarking in a live experiment involving nearly twenty
million Gemini responses~\cite{Dathathri2024}. Golowich and Moitra make the coding
connection explicit by constructing pseudorandom codes that tolerate insertions,
deletions, and substitutions and using them to obtain text watermarks with provable
robustness to editing under their stated assumptions~\cite{GM2024}. These applications
require information to remain recoverable or recognizable even when the sequence
carrying it has changed.

The origins of the subject lie in the study of communication with synchronization errors.
Sellers's work on bit loss and gain appeared in 1962, followed by Levenshtein's
foundational combinatorial treatment in the mid-1960s. An early discovery connected this
new problem to a construction designed for a different purpose. Varshamov and Tenengolts
had introduced codes for correcting a single asymmetric error, and Levenshtein recognized
that their weighted-checksum construction also corrects a single deletion. These codes
achieve essentially the minimum possible redundancy for that task. The single-deletion
acase thus offered an unusually encouraging starting point, with a compact construction, an
efficient recovery procedure, and a guarantee close to the information-theoretic
limit~\cite{Sellers1962,VT65,Levenshtein1966}.

Yet extending this early success to multiple deletions proved far more difficult. Helberg
codes offered an important generalization of the weighted-checksum approach, proposed by
Helberg and Ferreira and later given a complete proof by Abdel-Ghaffar and colleagues.
These codes correct multiple deletions, but at a substantial cost. Even for two deletions,
their redundancy grows linearly in the length of the transmitted sequence. The
construction thus extended the reach of the checksum method while losing the remarkable
economy of its single-deletion counterpart. Obtaining explicit codes that combined
efficient encoding and decoding with similarly small redundancy remained a central
challenge for decades~\cite{HF2002,APFC2012,BGZ2018}.

The problem also prompted a broader information-theoretic question about how much
information a channel can carry when synchronization itself is unreliable. Dobrushin's
1967 coding theorem provided an early foundation for studying channels in which an input
symbol can generate a random output sequence, including no output at all. This stochastic
viewpoint asks about reliable transmission under a specified random error process. The
combinatorial viewpoint instead demands recovery from every error pattern within a
prescribed budget. Both study the loss of alignment, but their guarantees are different,
with high-probability recovery over a random channel in one case and protection against
the worst allowed pattern in the other~\cite{Dobrushin1967}.

The algorithmic side of synchronization has since become an important meeting point of
coding theory and theoretical computer science. In 1999, Schulman and Zuckerman
constructed codes with efficient encoding and decoding that protect a constant proportion
of information against a constant proportion of insertions and deletions. Later, Haeupler
and Shahrasbi introduced synchronization strings, a method of attaching robust positional
information to a sequence. By helping the receiver recover alignment, this method allows
techniques for ordinary symbol errors and erasures to be adapted to insertion/deletion
channels. These developments show how identifying positions can itself be treated as a
coding problem, rather than assumed to have been solved by the communication
system~\cite{SZ1999,HS2021}.

The connection with document exchange gives a complementary route to code construction. A
short summary that lets a receiver recover an original file from a nearby version can
serve as the redundancy of a code, provided the summary itself is protected appropriately.
Haeupler's work develops this connection through efficient document-exchange protocols and
derives systematic insertion/deletion-correcting codes from them. The relationship makes
progress on communication-efficient reconciliation relevant to the design of codes for
synchronization errors~\cite{Haeupler2019}.

Within this broader history, correcting a fixed number of deletions poses a particularly
sharp challenge. The message grows longer while the number of errors stays small, so
almost all of the transmitted symbols survive. The aim is to protect those messages with a
vanishing fraction of redundant bits, while preserving efficient encoding and decoding.
The classical single-deletion construction achieves precisely this kind of economy.
Multiple deletions are harder to separate because several losses of alignment can
interact, and a check that resolves one missing symbol need not distinguish all the
possibilities created by several.

Brakensiek, Guruswami, and Zbarsky made a major advance by constructing efficient binary
codes for any fixed number of deletions with the same qualitative dependence of redundancy
on message length as in the single-deletion case. Multiple deletions could therefore be
handled efficiently without a fundamentally larger growth in overhead, although
substantial room remained to reduce the amount of redundancy. For two deletions, Guruswami
and H{\aa}stad subsequently obtained explicit codes matching the leading redundancy of the
classical greedy existence guarantee. Their construction builds on the single-deletion
checksum by adding checks that distinguish the more complicated ambiguities created by a
second deletion. These results illustrate the central algorithmic challenge. Counting
arguments can show that enough well-separated messages exist without revealing how to
encode messages into such a set or recover them efficiently~\cite{BGZ2018,GH2021}.

For a fixed number $t$ of errors, the classical Gilbert--Varshamov-type counting argument
guarantees binary codes with redundancy at most $2t\log_2 n+O_t(1)$, even when the errors
may be any combination of insertions, deletions, and substitutions. The direct greedy
construction, however, searches an exponentially large space.

Recent results highlight the remaining algorithmic challenge. Bui, Wang, and Vu give
deterministic polynomial-time encoding and decoding for binary codes correcting $t$ such
edits with redundancy $(2t+1)\log_2 n+O_t(\log\log n)$. On the existential side, En Gad
obtains the smaller redundancy $(2t-1)\log_2 n+O_t(\log\log n)$ for fixed $t\ge2$. The
classical Gilbert--Varshamov guarantee is therefore an algorithmic benchmark rather than
the strongest known existence bound~\cite{BWV2026,EG2026}.

Our focus is to turn the classical Gilbert--Varshamov guarantee into an efficient code
construction. We ask the following question.

\begin{problem}
For every fixed $t\ge2$, can one construct binary codes that correct any combination of at
most $t$ insertions, deletions, and substitutions in total, with redundancy at most
$2t\log_2 n+O_t(1)$, using deterministic algorithms for construction, encoding, and
decoding that run in polynomial time in $n$?
\end{problem}

We answer this question affirmatively. To the best of our knowledge, this is the \textit{first} binary code construction to attain the classical Gilbert--Varshamov guarantee up to an additive constant for every fixed $t\ge2$, while supporting deterministic polynomial-time construction, encoding, and
decoding. In particular, our result removes the additional $\log_2 n$ term in the redundancy
bound of Bui, Wang, and Vu~\cite{BWV2026} and achieves constant additive overhead
above $2t\log_2 n$.

Moreover, it provides a surprisingly simple solution to this longstanding
construction problem. As we explain in the next section, the proof uses elementary
counting and linear algebra. It combines counting solutions to affine binary
systems, applying the method of conditional expectations, and encoding messages
one bit at a time using additive lower bounds on the number of valid completions.
These ideas also apply beyond the edit channel. Section~\ref{sec:proof-ideas}
compares them with previous approaches and discusses extensions to more general
error types.

This paper is organized as follows. Section~\ref{sec:our_result} states the main result, reviews the
Gilbert--Varshamov-type bound, and outlines the construction.
Section~\ref{sec:preliminaries} develops the affine representation of
edit patterns and the counting procedure used throughout the proof.
Section~\ref{sec:construction} gives the deterministic construction,
encoding, and decoding algorithms and proves the main theorem.
Section~\ref{sec:example} illustrates the method with a worked example.
Section~\ref{sec:conclusion} concludes with directions for further work.

\section{Our Result: From Random Linear Hashing to An Efficient Code Construction} 
\label{sec:our_result}

\subsection{Error model and the Gilbert--Varshamov bound}\label{sec:error-model}

We focus on the adversarial worst-case regime with a fixed number
of errors $t\ge1$, independent of the block length $n$. An \textbf{\emph{edit}}
is an insertion, deletion, or substitution of a single bit. For
$x\in\{0,1\}^n$, let $B_t(x)$ denote its \textbf{\emph{error ball}},
consisting of all received words that can arise from $x$ after
up to $t$ edits. A binary \textbf{\emph{code}} $C\subseteq\{0,1\}^n$
corrects $t$ edits if the error balls associated with distinct
codewords are pairwise disjoint. Its \emph{redundancy} is
$r(C)=n-\log_2|C|$.

Two distinct words $x,x'\in\{0,1\}^n$ are called
\textbf{\emph{confusable}}, or \textbf{\emph{confusable neighbors}}, if
\[
 B_t(x)\cap B_t(x')\ne\varnothing,
\]
meaning that they can produce the same received word. A code corrects $t$ edits
precisely when no two of its codewords are confusable.

Let $\Delta$ be the maximum number of confusable neighbors of any word. A greedy
construction repeatedly selects a remaining word as a codeword and removes it
together with its confusable neighbors. Each selection removes at most $\Delta+1$
words, so the resulting code $C$ satisfies
\[
 |C|\ge \frac{2^n}{\Delta+1},
 \qquad
 r(C)\le \log_2(\Delta+1).
\]
This Gilbert--Varshamov-type argument gives an achievable
redundancy bound, but does not itself provide efficient
encoding and decoding algorithms for a code attaining
this bound.

Two distinct length-$n$ words are confusable precisely when
their edit distance is at most $2t$. Counting sequences of at
most $2t$ edits gives
\[
 \Delta=O_t(n^{2t}).
\]
The greedy argument therefore guarantees a binary
$t$-edit-correcting code with redundancy at most
$2t\log_2 n+O_t(1)$. Although this bound guarantees the code
size we seek to achieve, a direct implementation of the greedy
construction requires enumerating $\F_2^n$ and therefore takes
exponential time in $n$.

\subsection{From random hashing to an efficient construction}\label{sec:construction-overview}

In order to overcome this difficulty, we begin by first organizing candidate words 
by their syndromes under a linear transformation. Specifically, a binary matrix
$H\in\F_2^{b\times n}$ partitions the ambient space into syndrome
classes, or cosets,
\[
 D_{H,\beta}=\{x\in\F_2^n:Hx=\beta\}.
\]
Call a word \textbf{\emph{separated}} if none of its distinct confusable neighbors
has the same syndrome.
If every word in $D_{H,\beta}$ is separated, then the coset itself is a
$t$-edit-correcting code, described simply by the equations $Hx=\beta$.

Unfortunately, this simple description does not necessarily
yield high-rate codes. For instance, when $\beta=0$, the coset is a
linear code, and correcting even a single deletion limits its rate to
at most $1/2$. We instead settle for the much weaker requirement of
finding a coset in which at least half the words are separated. As we
show next, random hashing guarantees the existence of such a coset
whose separated words form a code meeting the GV redundancy target,
with at most one additional bit of redundancy relative to the full coset.

Let $H$ denote a uniformly random matrix and let
$x,x'$ be distinct vectors. Under this setting, 
$\Pr[Hx=Hx']=2^{-b}$, so a union bound gives
\[
 \Pr[x\text{ is not separated}]\le\Delta\,2^{-b}\le\tfrac12
 \qquad\text{when }2^b\ge2\Delta.
\]

Averaging
over the $2^b$ syndrome classes gives a coset with at least
$2^{n-b-1}$ separated words. These words form a code with redundancy
at most $b+1$. Taking $b=2t\log_2 n+O_t(1)$ to ensure $2^b\ge2\Delta$
therefore achieves the GV redundancy target up to an additive constant.

The challenge and aim of this work is to realize this size guarantee \emph{efficiently}. We accomplish this by way of the following three part roadmap:

\begin{enumerate}
\item \textbf{Part~1: Construct the matrix \textcolor{BrickRed}{(Section~\ref{sec:linear-hashing})}.} Choose a full-row-rank matrix $H$
for which half of the words are separated.
\item \textbf{Part~2: Identify the coset \textcolor{BrickRed}{(Section~\ref{sec:syndrome-selection})}.} Find a coset $D_{H,\beta}$ and
certify at least $2^{n-b-1}$ separated words within it.
\item \textbf{Part~3: Index the codewords \textcolor{BrickRed}{(Section~\ref{sec:separated-encoding})}.} Map information messages injectively
to a guaranteed subset of these words and recover each message from its
codeword, in polynomial time for fixed $t$.
\end{enumerate}

All three parts can be implemented \textit{efficiently} using the same counting idea, which we develop in Section~\ref{sec:counting-procedure}. 

Combining Parts~1--3 gives the following result.
\begin{theorem}[Main result]\label{thm:main}
For every fixed integer $t\ge1$ and all sufficiently large $n$, there is a
binary code $\mathcal C\subseteq\F_2^n$ that corrects any combination of
at most $t$ insertions, deletions, and substitutions and has redundancy
\[
 r(C):= n-\log_2|\mathcal C|\le 2t\log_2 n+O_t(1).
\]
The shared code parameters can be constructed deterministically in time
polynomial in $n$. Encoding and decoding, including recovery of the
information message, are also deterministic and take time polynomial
in $n$ for fixed $t$.
\end{theorem}

\subsection{Comparison with prior constructions and extensions}\label{sec:proof-ideas}

We now compare our construction with earlier algorithmic approaches. We begin
with the recent work of Bui, Wang, and Vu, explaining how the constructions differ
and why these differences lead to a simpler construction with smaller redundancy.
We then relate our approach to syndrome compression, document exchange, and graph
coloring, before discussing its extension to other error models.

Bui, Wang, and Vu~\cite{BWV2026} use a technique they refer to as polynomial hashing to distinguish each payload
from its confusable alternatives. Their encoder chooses a suitable hash for each
payload and appends a protected tag identifying both the chosen hash and its
value. Their entropy-recycling step uses the choice among suitable hashes to
carry additional message bits, substantially reducing the redundancy of this
representation.

Our construction changes how hashing is used. One binary matrix and one syndrome
are fixed for the entire code. We encode messages directly into the separated
words of that class, so no codeword needs to carry a payload-dependent seed or
hash tag. The choice of the shared parameters is made once during code
construction. Encoding then selects codewords using the same counting procedure.
This removes the separate seed-selection, tag-protection, and entropy-recycling
components from the encoder.

The hash family also matters. Once its seed is fixed, the polynomial hash of
Bui, Wang, and Vu is itself binary linear. The distinction is therefore in the
collision guarantee. Their bound includes a factor depending on the polynomial
degree, which grows with the payload length. A uniformly random binary matrix
has no such factor, allowing a shorter hash. Direct encoding into a fixed class
also eliminates the overhead of protecting the tag. Together, these changes
reduce the redundancy from $(2t+1)\log_2 n+O_t(\log\log n)$ to
$2t\log_2 n+O_t(1)$. Both constructions support deterministic polynomial-time
encoding and decoding for fixed $t$. Bui, Wang, and Vu additionally preserve most
message bits unchanged in the codeword, a property our construction does not
require~\cite{BWV2026}.

Our construction also differs from earlier approaches based on synchronization
structure and compressed checks. Sima and
Bruck~\cite{SB2021} use synchronization patterns to divide a word into short blocks
and protect their boundaries. Sima, Gabrys, and Bruck~\cite{SGB2020} instead use
successive indicator representations and compress the resulting checks. Their
general syndrome-compression method starts with a label that distinguishes each
input from its competitors, then shortens it using a modulus chosen for that
input. Both the modulus and the reduced label must be specified~\cite{SGBSC2020}.
Our construction works directly with affine descriptions of the errors, avoiding
the need to design and compress a separating label for each input.

Document exchange highlights another useful distinction. A short sketch must
allow the receiver to reconstruct an arbitrary input from a nearby
word~\cite{CJLW2022,Haeupler2019}. Our encoder has the freedom to map messages into
a selected subset of words. The shared hash therefore need only separate enough
words to support the desired number of messages. Efficiently identifying and
encoding into that subset lets us exploit this freedom.

The requirement that a document-exchange sketch distinguish confusable inputs has a natural graph interpretation. Li, Gabrys, and Farnoud~\cite{LGF2026}
use distributed graph coloring to assign different labels to every pair of
confusable words. Their method computes the label of an individual word
efficiently while remaining consistent with a proper coloring of the whole
graph. Our hash may leave some confusable words with the same label. We use only
the separated words in one class and encode directly into that set. Their
framework applies to more general confusion graphs, whereas our efficient
implementation relies on a polynomial-size family of affine maps covering all
confusable pairs. The size of such a family need not equal the graph's maximum
degree.

This reliance on affine maps also identifies when the construction extends to
other error models. The maps must cover every confusable pair and be enumerable
in polynomial time. Under these conditions, the same argument gives deterministic
polynomial-time construction and encoding for sufficiently large block lengths,
with redundancy at most the base-two logarithm of the family size plus a constant.
Efficient decoding additionally
requires a polynomial-time procedure for enumerating the possible transmitted
words consistent with a received word. The fixed syndrome then identifies the
unique candidate for any valid transmission.

Adjacent transpositions provide a concrete example. Each fixed sequence of swaps
is a permutation of coordinates and hence a linear map. If at most $t$ adjacent
swaps are allowed, two confusable words differ by at most twice that number of
swaps. For fixed $t$, polynomially many such maps cover all confusable pairs, and the
possible transmitted words can be enumerated by reversing the swaps. The method
therefore gives the same redundancy guarantee as Theorem~\ref{thm:main} for this
error model, with deterministic polynomial-time encoding and decoding.

We now develop the affine representation of edit patterns and the counting
procedure that make these constructions possible.

\section{Preliminaries: Affine Error Patterns and A Counting Procedure}\label{sec:preliminaries}

For a word $x$, its length is denoted by $|x|$, and $x_{a:b}$ denotes the substring from position $a$ through position $b$. The empty word
is denoted by $\varepsilon$. In particular, $x_{1:0}=\varepsilon$.
For a word $p$ and a bit $a$, $pa$ denotes their concatenation.
We write $\langle u,v\rangle$ for the binary inner product.

\subsection{Affine transformations}\label{sec:affine-model}

A map $f:\F_2^n\to\F_2^n$ is \emph{affine} if $f(x)=Ax+e$ for a
fixed matrix $A$ and vector $e$ over $\F_2$.

Fix an error pattern $\sigma$ that maps length-$n$ binary words to
length-$n$ binary words. The error pattern specifies the deleted positions,
the positions and values of inserted bits, and the positions of
substitutions. Each output coordinate is then a copied input bit, a
complemented input bit, or a constant. Thus, over $\F_2$, its action has
the form
\[
 f_\sigma(x)=A_\sigma x+e_\sigma,
\]
where the output is a function of $x$. Define the difference vector
\[
 d_\sigma(x)=x+f_\sigma(x)=(I+A_\sigma)x+e_\sigma.
\]

\begin{example}\label{ex:affine_map}
Let $n=4$, and let $\sigma$ delete the second input bit and append a one.
Then
\[
 \begin{aligned}
 f_\sigma(x)&=(x_1,x_3,x_4,1),\\
 d_\sigma(x)&=(0,x_2+x_3,x_3+x_4,x_4+1).
 \end{aligned}
\]
Here
\[
 A_\sigma=
 \begin{pmatrix}
 1&0&0&0\\
 0&0&1&0\\
 0&0&0&1\\
 0&0&0&0
 \end{pmatrix},
 \qquad
 e_\sigma=
 \begin{pmatrix}
 0\\0\\0\\1
 \end{pmatrix}.
\]
The first three rows of $A_\sigma$ retain input coordinates $1,3,4$,
while $e_\sigma$ supplies the appended one.
For instance, $x=0100$ gives $f_\sigma(x)=0001$ and
$d_\sigma(x)=0101$. The same error pattern leaves $0111$ unchanged, so its difference vector is zero.
\end{example}

In this setting, let $\Sigma$ index a family of $M$ affine
error maps induced by length-preserving error patterns of
at most $2t$ edits and covering all confusable neighbors.
We now develop a counting procedure that, for each
$\sigma\in\Sigma$, uses the affine representation of
$f_\sigma$ to efficiently count words satisfying prescribed
affine constraints, without explicitly listing or testing
those words individually. 

\subsection{The central counting procedure}\label{sec:counting-procedure}

For a fixed matrix $H$, a \emph{collision witness} is an input--error pattern pair
$(x,\sigma)$ such that $f_\sigma(x)$ is distinct from $x$ but has the same syndrome, i.e., 
\[
 d_\sigma(x)\ne0,\qquad Hd_\sigma(x)=0.
\]

For each word $x$, let
\begin{align}
W_H(x)&=\big\vert\{\sigma\in\Sigma:
d_\sigma(x)\ne0,\ Hd_\sigma(x)=0\}\big\vert \\
&\geq \mathbf1_{\{x\text{ is not separated under }H\}}
\end{align}
be its number of collision witnesses.  Observe that $W_H(x)=0$ if and only if $x$ is separated.
For a set of candidate words $S\subseteq\F_2^n$, we define the number of collision witnesses under the matrix $H$ w.r.t. the candidate set $S$ as follows: 
\begin{align}\label{eq:WHS}
 W_H(S)&:=\sum_{x\in S}W_H(x)\\ 
 &\geq \big\vert\{x\in S:x\text{ is not separated under }H\}\big\vert.
\end{align}

Grouping the collision witnesses in~\eqref{eq:WHS} by error pattern gives
\begin{equation}\label{eq:affine-count}
 W_H(S)=\sum_{\sigma\in\Sigma}\Bigl(
 \bigl|\{x\in S:Hd_\sigma(x)=0\}\bigr|
 -\bigl|\{x\in S:d_\sigma(x)=0\}\bigr|\Bigr).
\end{equation}
For each $\sigma$, the first term counts inputs in $S$ whose images have
the same syndrome, including fixed points. The second term removes
those fixed points, since $d_\sigma(x)=0$ implies $Hd_\sigma(x)=0$.

If $S$ is specified by affine equations, each count
in~\eqref{eq:affine-count} is the number of solutions of an affine
binary system. For a system $Cx=a$ in $n$ unknowns, the number of solutions is
\[
 \big\vert\{x:Cx=a\}\big\vert 
 =\begin{cases}
 2^{n-\operatorname{rank}C},
 &\operatorname{rank}C=\operatorname{rank}[C\mid a],\\
 0,&\operatorname{rank}C<\operatorname{rank}[C\mid a].
 \end{cases}
\]
Gaussian elimination determines consistency and rank.

\begin{excont}{ex:affine_map}
We illustrate the computation of $W_H(S)$ in~\eqref{eq:affine-count}
using the family consisting only of the preceding error map, so
$\Sigma=\{\sigma\}$ within this example. Its difference vector is
\[
 d_\sigma(x)=(0,x_2+x_3,x_3+x_4,x_4+1).
\]
For illustration, choose $H=(0,1,0,0)$. Since $Hx=x_2$, the
syndrome class $S=D_{H,1}$ consists exactly of the words with $x_2=1$.
We first count words in $S$ whose images under the error map have the same syndrome.
Membership in $S$ gives $x_2=1$, while $Hd_\sigma(x)=0$ gives
$x_2+x_3=0$. These are the two rows of the system $Cx=a$:
\[
 \underbrace{\begin{bmatrix}
 0&1&0&0\\
 0&1&1&0
 \end{bmatrix}}_{C}x
 =\underbrace{\begin{bmatrix}1\\0\end{bmatrix}}_{a}.
\]
Here $\operatorname{rank}C=\operatorname{rank}[C\mid a]=2$,
so the first count is $2^{4-2}=4$. Indeed, the equations force
$x_2=x_3=1$, leaving $x_1$ and $x_4$ free.

This count includes words left unchanged by the error map.
To count these fixed points, we impose $d_\sigma(x)=0$ in place of
$Hd_\sigma(x)=0$. In addition to $x_2+x_3=0$, this requires
$x_3+x_4=0$ and $x_4+1=0$; the first coordinate of $d_\sigma(x)$
is already zero. Together with $x_2=1$, these equations give
\[
 \underbrace{\begin{bmatrix}
 0&1&0&0\\
 0&1&1&0\\
 0&0&1&1\\
 0&0&0&1
 \end{bmatrix}}_{C}x
 =\underbrace{\begin{bmatrix}1\\0\\0\\1\end{bmatrix}}_{a}.
\]
Now $\operatorname{rank}C=\operatorname{rank}[C\mid a]=3$,
so the fixed-point count is $2^{4-3}=2$: the equations force
$x_2=x_3=x_4=1$, leaving only $x_1$ free.
The two cardinalities in~\eqref{eq:affine-count} are therefore $4$ and $2$, giving
\[
 W_H(S)=4-2=2.
\]
Thus we obtain two collision witnesses without enumerating the words in $S$.
\end{excont}

For each $\sigma\in\Sigma$, one affine-system count evaluates
$\bigl|\{x\in S:Hd_\sigma(x)=0\}\bigr|$, and a second evaluates
$\bigl|\{x\in S:d_\sigma(x)=0\}\bigr|$.
Thus $2M$ such counts suffice to evaluate $W_H(S)$ exactly.
When $S$ is given by polynomially many affine equations, this takes
polynomial time in $n$ and $M$, without enumerating $S$.

Next, we define a \emph{signed lower bound} on the number of separated words under the matrix $H$ w.r.t. the candidate set $S$ as follows: 
\begin{equation}\label{eq:bound}
 L_H(S):=\sum_{x\in S}\bigl(1-W_H(x)\bigr) =  |S| - W_H(S).
\end{equation}
A separated word contributes $1$, whereas a word with $k\ge1$
collision witnesses contributes $1-k\le0$. Therefore
\[
 \mathbf1_{\{x\text{ is separated under }H\}}\geq 1-W_H(x),
\]
and summing over $S$ yields
\begin{equation}\label{eq:lower}
 \big\vert\{x\in S:x\text{ is separated under }H\}\big\vert\ge L_H(S).
\end{equation}

For fixed $H$, the collision witness count $W_H(x)$ depends only on $x$,
not on the candidate set containing it. Hence, for disjoint sets
$S_0,S_1$,
\begin{align}\label{eq:additivity}
 W_H(S_0\cup S_1)&=W_H(S_0)+W_H(S_1).\\
 L_H(S_0\cup S_1)&=L_H(S_0)+L_H(S_1)\;.
\end{align}
This identity holds even when one of the terms is negative. This will later
allow us to distribute the bound on the whole space among syndrome classes and,
within a selected syndrome class, among prefix extensions.

The construction applies this additive signed lower bound as follows.
\begin{enumerate}
 \item \textbf{Choose the matrix.} Construct $H$ with
 at most $2^{n-1}$ surviving collision witnesses, i.e. $W_H(\F_2^n) \leq 2^{n-1}$, guaranteeing that
 at least half of all words are separated.
 \item \textbf{Choose the syndrome.} Partition $\F_2^n$ into syndrome
 classes and find a class $D_{H,\beta}$ with
 $W_H(D_{H,\beta})\le2^{n-b-1}$, guaranteeing that
 at least half of all words are separated.
 \item \textbf{Index the codewords.} Restrict the selected class by
 successive prefixes and use the additive signed bounds to encode
 $n-b-1$ message bits.
\end{enumerate}

\section{The construction}\label{sec:construction}

We now prove Theorem~\ref{thm:main} by carrying out the three parts
outlined above. We then combine their guarantees to establish the
redundancy bound and efficient encoding/decoding.

\subsection{Part 1: Choosing \texorpdfstring{$H$}{H}}\label{sec:linear-hashing}

We now proceed with the first step: choosing $H$. We construct it using the method of conditional expectations \cite{AlonSpencer2016}, one
row at a time, choosing each row to separate at least half the collision
witnesses surviving the preceding rows.

A parity-check row $h$ separates $x$ from $f_\sigma(x)$ precisely when
\[
 \ip{h}{d_\sigma(x)}=1,
 \qquad\text{equivalently,}\qquad
 \ip{h}{x}\ne\ip{h}{f_\sigma(x)}.
\]

To choose the next row bit by bit, let $G$ contain the rows already
chosen. Let $T$ count the surviving collision witnesses under G, i.e., $T = W_G(\F_2^n)$.

\begin{samepage}
Fix the first $j$ bits of the next row to the next partially completed row, which we refer to as $p$, and complete the remaining
bits uniformly at random. For each surviving witness, there are two cases:
\begin{itemize}
\item \textbf{Nonzero difference suffix: survival probability $1/2$.}
If $d_\sigma(x)_{j+1:n}\ne0$, at least one remaining row bit controls
the parity, making zero and one equally likely. 
\item \textbf{Zero difference suffix: the prefix decides.} 
If $d_\sigma(x)_{j+1:n}=0$, the remaining row bits have no effect.
The witness survives exactly when $\ip{p}{d_\sigma(x)_{1:j}}=0$. 
\end{itemize}
\end{samepage}
Let $Z_j$ count the witnesses with zero difference suffix and $Q_j(p)$
count those among them whose prefix parity is zero. 

For each $\sigma \in \Sigma$, let
\begin{align*}
S_{\sigma,j} &:= \{x \in \F_2^n : d_\sigma(x)_{j+1:n}=0\}\\
R_{\sigma,j}(p) &:= \{x \in \F_2^n :d_\sigma(x)_{j+1:n}=0, \ip{p}{d_\sigma(x)_{1:j}}=0\}\;.
\end{align*}
Then, it follows that
\[
 Z_j=\sum_{\sigma\in\Sigma}\Bigl(
 \bigl|\{x\in S_{\sigma,j}:Gd_\sigma(x)=0\}\bigr|
 -\bigl|\{x\in S_{\sigma,j}:d_\sigma(x)=0\}\bigr|\Bigr),
\]
and
\[
 Q_j(p)=\sum_{\sigma\in\Sigma}\Bigl(
 \bigl|\{x\in R_{\sigma,j}(p):Gd_\sigma(x)=0\}\bigr|
 -\bigl|\{x\in R_{\sigma,j}(p):d_\sigma(x)=0\}\bigr|\Bigr).
\]

Conditioned on the previously chosen rows $G$ and the prefix $p$
of the next row, the expected number of surviving witnesses is
\begin{equation}\label{eq:conditional-count}
 E(p)=\frac{T-Z_j}{2}+Q_j(p).
\end{equation}

\begin{lemma}[Counting under a partial row]\label{lem:count}
We can compute $E(p)$ exactly and construct a row separating at least
half the $T$ surviving witnesses, both in deterministic polynomial time.
\end{lemma}
\begin{proof}
First compute $T=W_G(\F_2^n)$ using the counting procedure from
Section~\ref{sec:counting-procedure}, which requires at most $2M$
affine-system counts. 
For each prefix $p\in\F_2^j$, at most $4M$ additional affine-system
counts suffice to evaluate $Z_j$ and $Q_j(p)$, and hence $E(p)$.
Specifically, for each error pattern $\sigma\in\Sigma$, two counts evaluate its
contribution to $Z_j$ using $S_{\sigma,j}$, and two more evaluate its
contribution to $Q_j(p)$ using $R_{\sigma,j}(p)$.
Since these sets are specified by polynomially many
affine equations, each prefix evaluation takes time polynomial in $n$
and $M$.

For $|p|<n$, a uniformly random next bit gives
\[
 E(p)=\frac{E(p0)+E(p1)}2.
\]
Consequently,
\[
 \min_{a\in\{0,1\}}E(pa)
 =E(p)-\frac{|E(p0)-E(p1)|}{2}
 \le E(p).
\]
Computing both values and appending a minimizing bit never increases
the expectation. Since $E(\varepsilon)=T/2$, the completed row $h$
satisfies $E(h)\le T/2$. Here $E(h)$ is the actual survivor count,
so at least half the witnesses are separated.
\end{proof}

\paragraph*{Completing the matrix}
Let $M=|\Sigma|\ge1$ and set $b=\lceil\log_2(2M)\rceil$, with $b<n$.
Initially there are at most $M2^n$ witnesses. Apply Lemma~\ref{lem:count}
successively to choose $b$ rows. While witnesses remain, each new row
is linearly independent of the preceding rows since a row in their span would
separate no further witnesses. If no witnesses remain, extend the matrix to rank $b$
with independent rows. Adding rows cannot create a collision. Thus
$\operatorname{rank}H=b$ and the total number of surviving witnesses is
\begin{equation}\label{eq:global-witness-bound}
 \begin{aligned}
 W_H(\F_2^n)&\le M2^{n-b}\le2^{n-1}.
 \end{aligned}
\end{equation}

We have now constructed $H$ so that at least half of all words are
separated. Part~2 locates a syndrome class containing at least
$2^{n-b-1}$ separated words.

\subsection{Part 2: Choosing the syndrome}\label{sec:syndrome-selection}

Once $H$ is fixed, we choose a target syndrome $\beta$ whose class has
a large guaranteed number of separated words. This is a one-time setup
step and every encoded word will satisfy $Hx=\beta$. We try all syndromes
and choose the one with the smallest upper bound on the number of words that are not separated.

For a syndrome $\beta\in\F_2^b$, the number of words in its class is
\[
 |D_{H,\beta}|=\big\vert\{x\in\F_2^n:Hx=\beta\}\big\vert
 =2^{n-\operatorname{rank}H}=2^{n-b}.
\]
Every syndrome is attainable because $H$ has full row rank.

\begin{lemma}[Choosing a syndrome]\label{lem:syndrome}
We can find a syndrome $\beta^\star$ with $W_H(D_{H,\beta^*})\le2^{n-b-1}$
in deterministic polynomial time.
\end{lemma}
\begin{proof}

Since the candidate set $D_{H,\beta}$ is specified by polynomially many affine equations, the number of collision witnesses under $H$ w.r.t. the candidate set $D_{H,\beta}$, $W_H(D_{H,\beta})$, can be computed in polynomial time. 

We choose
\[
\beta^\star\in\arg\min_{\beta\in\F_2^b}W_H(D_{H,\beta}) 
\]

Since $\{D_{H,\beta}\}_{\beta \in \F_2^b}$  are disjoint cosets, using \eqref{eq:global-witness-bound} from Part~1, gives

\[
\begin{aligned}
\sum_{\beta\in\F_2^b}W_H(D_{H,\beta})
&=\sum_{\beta\in\F_2^b}\;\sum_{x\in D_{H,\beta}}W_H(x)\\
&=\sum_{x\in\F_2^n}W_H(x)\\
&=W_H(\F_2^n)\le 2^{n-1}.
\end{aligned}
\]
The second equality holds because each word belongs to exactly one
syndrome class.

There are $2^b$ syndromes, so the minimizing choice satisfies
$W_H(D_{H,\beta^*})\le2^{n-b-1}$. The exhaustive search over the $2^b$ choices for the syndromes is possible in polynomial time because there are
\[
 2^b=2^{\lceil\log_2(2M)\rceil}<4M.
\]
syndromes.
\end{proof}

Since the chosen class has $2^{n-b}$ words, at least half are certified
separated. We have now fixed the shared parameters $H$ and $\beta^\star$
and certified enough separated words for $n-b-1$ message bits.
Part~3 will index a subset of these words to encode messages efficiently
and recover each message from its codeword.

\subsection{Part 3: Indexing the codewords}\label{sec:separated-encoding}

Our encoder follows the classical enumerative-coding approach, which
indexes codewords through their prefix-completion
counts~\cite{Cover1973,OR2000,OR2007,OR2009}. In our setting, affine
collision-witness counts yield efficiently computable signed lower bounds
that are additive over the binary prefix tree. For completeness, we give
a self-contained proof of the encoder and its inverse on the encoder
image, working directly with these bounds.

Fix the matrix $H$ and write $\beta=\beta^\star$ for the syndrome chosen
in Part~2. The encoder builds a codeword one bit at a time. At each step, it uses lower bounds on the number of valid completions to determine how many messages to assign to each possible next bit. It never needs to list all the separated words.

For a prefix $p \in \F_2^{\leq n}$, let  the set of the set of words in $D_{H,\beta}$ with prefix $p$ be denoted by $D_{H,\beta}(p)$, i.e.,
\[
D_{H,\beta}(p) := \{x:\in\F_2^n, Hx = \beta, x_{1:|p|} = p\}\;.
\]
 By convention, $D_{H,\beta} := D_{H,\beta}(\varepsilon)$, where $\varepsilon$ is the empty word. 

Since the candidate set $D_{H,\beta}(p)$ is specified by polynomially many affine equations, it's cardinality $\vert D_{H,\beta}(p)\vert$ and the number of collision witnesses under $H$ w.r.t. the candidate set $D_{H,\beta}(p)$, $W_H(D_{H,\beta}(p))$, can be computed in polynomial time. Furthermore, since $L_H(D_{H,\beta}(p)) = |D_{H,\beta}(p)| - W_H(D_{H,\beta}(p))$, the signed lower bound $L_H(D_{H,\beta}(p))$ can also be computed in polynomial time. 

Although $L_H(D_{H,\beta}(p))$ may be negative, it is additive and partitioning the candidate set by the next bit of the prefix gives
\begin{equation}\label{eq:prefix-additivity}
 L_H(D_{H,\beta}(p)) = L_H(D_{H,\beta}(p0))+L_H(D_{H,\beta}(p1))\qquad (|p|<n).
\end{equation}

At a complete word $x$, the value of $L_H(D_{H,\beta}(x))$ is $1-W_H(x)$ if $Hx=\beta$, and zero
otherwise. Consequently, $L_H(D_{H,\beta}(x))>0$ exactly when $x$ is separated and has
syndrome $\beta$; in that case $L_H(D_{H,\beta}(x))=1$.

Set $K=n-b-1$ and interpret the $K$-bit message as an integer
$m\in\{0,\ldots,2^K-1\}$. Part~2 guarantees
$L_H(D_{H,\beta}(\varepsilon)):=L_H(D_{H,\beta}) := L_H(D_{H,\beta^*})\ge2^K$, so there are enough certified
separated words for all messages. Starting with the empty prefix
$p=\varepsilon$, construct the codeword one bit at a time, using $m$
as the current index.

At each step, compute $L_H(D_{H,\beta}(p0))$, the lower bound for separated words
extending the prefix $p0$.
\begin{itemize}
\item If $m<L_H(D_{H,\beta}(p0))$, append $0$ to $p$.
\item Otherwise, append $1$ and replace $m$ by
\[
 m-\max\{0, L_H(D_{H,\beta}(p0))\},
\] using the bound computed before appending the bit.
\end{itemize}

Indices below a positive $L_H(D_{H,\beta}(p0))$ use the zero branch. The remaining
indices use the one branch, subtracting $L_H(D_{H,\beta}(p0))$
from the current index so that it indexes the remaining choices within the one branch.
If $L_H(D_{H,\beta}(p0))\le0$, all indices use the one branch without changing the current index. Repeat until $|p|=n$.

\paragraph*{Example}
Suppose $m=7$, $L_H(D_{H,\beta}(p0))=6$, and $L_H(D_{H,\beta}(p1))=4$. Indices $0,\ldots,5$ use
the zero branch. Our index takes the one branch and becomes $m=1$:
it is now the second index in that branch. If instead $L_H(D_{H,\beta}(p0))=-2$
and $L_H(D_{H,\beta}(p1))=12$, we take the one branch and leave $m=7$ unchanged.

\begin{lemma}[Encoding and recovering messages]\label{lem:encoding}
The construction maps $K$-bit messages injectively to separated words
with syndrome $\beta$. Encoding and inversion on its image both take
time polynomial in $n$ and $M$.
\end{lemma}
\begin{proof}
We maintain $0\le m<L_H(D_{H,\beta}(p))$ along the selected path. This holds initially
by Part~2. On the zero branch it follows directly from $m<L_H(D_{H,\beta}(p0))$.
On the one branch, the new index is nonnegative and satisfies
\[
 \begin{aligned}
 &m-\max\{0,L_H(D_{H,\beta}(p0))\}\\
 &<L_H(D_{H,\beta}(p))-\max\{0,L_H(D_{H,\beta}(p0))\}\\
 &\le L_H(D_{H,\beta}(p))-L_H(D_{H,\beta}(p0))=L_H(D_{H,\beta}(p1)),
 \end{aligned}
\]
where the final equality is~\eqref{eq:prefix-additivity}.

At a complete word $x$, we have $0\le m<L_H(D_{H,\beta}(x))\le1$, forcing
$m=0$ and $L_H(D_{H,\beta}(x))=1$. The output is therefore separated and has syndrome
$\beta$. 

To prove injectivity, suppose two distinct initial message indices
$m_0\ne m'_0$ produce the same codeword $x$. Let $m_j$ and $m'_j$
denote their current indices after the first $j$ bits have been chosen.
Both runs follow the same sequence of prefixes, so each step subtracts
the same quantity from both indices, possibly zero. Hence
\[
 m_j-m'_j=m_0-m'_0\ne0
 \qquad\text{for every }j.
\]
At termination, both runs reach $x$. Each final index is a nonnegative
integer strictly less than $L_H(D_{H,\beta}(x))\le1$, so
$m_n=m'_n=0$. This gives
\[
 0=m_n-m'_n=m_0-m'_0\ne0,
\]
a contradiction. Therefore the encoder is injective.

To recover the message from a codeword in the image, start with
$p=\varepsilon$ and $m=0$, and read its bits from left to right.
Before appending each bit to $p$, add $\max\{0,L_H(D_{H,\beta}(p0))\}$ to $m$ if the
bit is $1$ and leave $m$ unchanged if it is $0$. These additions undo
the subtractions made during encoding. Since encoding ends with
index zero, the final sum is the original message index, written in
$K$ bits.

Each traversal evaluates $L_H(D_{H,\beta}(p0))$ at most $n$ times, and each evaluation can be computed in polynomial time. The running time is therefore polynomial in
$n$ and $M$, and hence in $n$ for fixed $t$.
\end{proof}

This encoding is generally nonsystematic: the message bits need not
appear in designated codeword positions.

\paragraph*{Proof of Theorem~\ref{thm:main}}
Lemma~\ref{lem:count} and the matrix completion argument in Part~1
construct a full-row-rank matrix $H$ with $b=\lceil\log_2(2M)\rceil$
rows and
\[
 W_H(\F_2^n)\le 2^{n-1}.
\]
Lemma~\ref{lem:syndrome} then finds a syndrome $\beta$ satisfying
\[
 L_H(D_{H,\beta})
 =2^{n-b}-W_H(D_{H,\beta})
 \ge 2^{n-b-1}.
\]
With these parameters, Lemma~\ref{lem:encoding} gives an injective
encoder for $K=n-b-1$ message bits, together with an efficient inverse
on its image. Taking this image as $\mathcal C$, we obtain
\[
 n-\log_2|\mathcal C|=b+1\le 2t\log_2 n+O_t(1),
\]
since $M=O_t(n^{2t})$. By Lemma~\ref{lem:encoding}, every codeword is
separated and has syndrome $\beta$. In other words, no two codewords are
confusable, and $\mathcal C$ corrects $t$ edits.

It remains to describe recovery from an edited word. To decode a
received word $y$ obtained from a codeword by at most $t$ edits,
enumerate the length-$n$ words within $t$ edits of $y$ and retain those
satisfying $Hx=\beta$. The transmitted word qualifies. Any distinct
retained word would be confusable with it and have the same syndrome,
contradicting separation. Thus this search uniquely recovers the
transmitted word, and the inverse from Lemma~\ref{lem:encoding}
recovers its message. The candidate search takes time polynomial in
$n$ for fixed $t$. Together with the deterministic polynomial-time
guarantees of Lemmas~\ref{lem:count}--\ref{lem:encoding}, this proves
Theorem~\ref{thm:main}.

\section{Example}\label{sec:example}
We illustrate the three parts of Section~\ref{sec:construction} with
$n=8$ and $t=1$. Take $\Sigma$ to consist of the $8$ single-bit
substitutions, the $\binom{8}{2}=28$ two-bit substitutions, and the
$2\cdot8^2=128$ deletion--insertion patterns. Thus $M=164$, and this
family covers all confusable neighbors for one edit.

Consider the full-row-rank matrix
\[
 H=\begin{pmatrix}
 0&0&1&0&0&1&1&0\\
 0&0&0&1&1&1&0&0\\
 0&1&0&0&0&0&1&0\\
 0&0&0&0&1&0&1&1\\
 1&0&0&0&0&1&0&1
 \end{pmatrix}.
\]

We verify that the above matrix separates more than $128$ words directly by grouping witnesses by their difference $d=x+x'$. Equal syndromes imply $Hd=0$. The kernel has dimension three, and its seven nonzero vectors are listed in the table below. All have Hamming weight at least four, so a collision cannot come from one or two substitutions. Only deletion--insertion patterns
remain.

Suppose a deletion--insertion pattern deletes $x_i$ and inserts a bit $a$
at position $j>i$ of the resulting word $x'$. Within positions $i$ through
$j$, the input and output blocks satisfy
\[
 (x'_i,\ldots,x'_{j-1},x'_j)
 =(x_{i+1},\ldots,x_j,a).
\]
We call this a left shift of the interval $[i,j]$. A right shift instead
deletes $x_j$ and inserts $a$ at position $i$, replacing the same input
block with $a x_i\cdots x_{j-1}$. Both operations leave every coordinate
outside $[i,j]$ unchanged.

Now prescribe the difference $d=x+x'$. Since the two words agree outside
$[i,j]$, this interval must contain every nonzero coordinate of $d$.
For a left shift, the differences within the interval satisfy
\[
 (d_i,\ldots,d_{j-1},d_j)
 =(x_i+x_{i+1},\ldots,x_{j-1}+x_j,x_j+a).
\]

Write $\ell=j-i+1$ for the interval length. There are two possible shift
directions, left and right. For each direction, the inserted bit $a$ can
be either $0$ or $1$. Each of these four choices uniquely determines the
bits inside the interval and allows $2^{8-\ell}$ assignments to the bits
outside it. Thus, the number of collision witnesses for this difference
and interval is
\[
 2\cdot2\cdot2^{8-\ell}=4\cdot2^{8-\ell}.
\]

For example, $d=00101101$ has its first and last nonzero coordinates
at positions $3$ and $8$. The possible intervals are $[3,8]$, $[2,8]$,
and $[1,8]$. Since the interval length is $\ell=j-i+1$, their
respective contributions are
\[
\begin{aligned}
{[3,8]}:&\qquad 4\cdot 2^{8-6}=16,\\
[2,8]:&\qquad 4\cdot 2^{8-7}=8,\\
[1,8]:&\qquad 4\cdot 2^{8-8}=4.
\end{aligned}
\]
Thus this difference contributes $16+8+4=28$ witnesses. The same calculation
for the other differences gives
\begin{center}
\begin{tabular}{ccc}
\hline
Difference $d$ & Interval lengths & Witnesses\\
\hline
$00101101$ & $6,7,8$ & $4(4+2+1)=28$\\
$01010111$ & $7,8$ & $4(2+1)=12$\\
$01111010$ & $6,7,7,8$ & $4(4+2+2+1)=36$\\
$10011001$ & $8$ & $4$\\
$10110100$ & $6,7,8$ & $4(4+2+1)=28$\\
$11001110$ & $7,8$ & $4(2+1)=12$\\
$11100011$ & $8$ & $4$\\
\hline
\end{tabular}
\end{center}
The two length-seven intervals in the third row are $[1,7]$ and
$[2,8]$. We count patterns separately, as required by $W_H$. Summing,
\[
 W_H(\F_2^8)=28+12+36+4+28+12+4=124\le128.
\]
This certifies that the above matrix separates more than $128$ words. For this
small example, $b=5$ rows suffice. The general procedure for choosing
rows is proved in Lemma~\ref{lem:count}.

Take the nonzero syndrome $\beta=(1,0,0,0,0)^T$. Its eight words are
listed below. The last column records the message assignments from Part~3.
\begin{center}
\begin{tabular}{ccc}
\hline
Word $x$ & Separated? & Message\\
\hline
$00001101$ & Yes & $00$\\
$00100000$ & Yes & $01$\\
$01011010$ & Yes & $10$\\
$01110111$ & No & ---\\
$10010100$ & Yes & $11$\\
$10111001$ & Yes & ---\\
$11000011$ & Yes & ---\\
$11101110$ & No & ---\\
\hline
\end{tabular}
\end{center}

We next determine which words in $D_{H,\beta}$ are separated and
evaluate the corresponding signed lower bound. Since all eight words
have syndrome $\beta$, a word is separated precisely when none of
the other seven words is confusable with it.

For distinct $x,x'\in D_{H,\beta}$, the difference $x+x'$ is a is one of the seven nonzero kernel vectors listed above, and therefore has Hamming weight at least four. Thus one or two substitutions cannot transform one word into another. The only remaining possibility for confusability is a deletion followed by an insertion. Equivalently, the two words must share a length-seven subsequence.

Comparing the one-deletion outputs of the eight words shows that the only confusable pair is $01110111$ and $11101110$. Their common output is
\[
\begin{aligned}
01110111&\xrightarrow{\text{delete bit }1}1110111,\\
11101110&\xrightarrow{\text{delete bit }8}1110111.
\end{aligned}
\]
This is their only common one-deletion output, and it is obtained
uniquely from each word. Hence there is exactly one collision witness
in each direction: deleting the initial zero and appending a zero
maps $01110111$ to $11101110$, while deleting the final zero and
prepending a zero gives the reverse map. The other six words have
no collision witnesses and are therefore separated.

Although there is only one unordered confusable pair, it contributes
two witnesses, one with each word as its input. Consequently,
\[
 W_H(D_{H,\beta})=2.
\]
Each of the six separated words contributes $1$ to the signed bound,
and each of the two remaining words contributes $1-1=0$. Therefore
\[
 L_H(D_{H,\beta})=6=8-2.
\]
In this class, the signed bound equals the actual number of separated
words. Since
\[
 L_H(D_{H,\beta})=6\ge 2^{n-b-1}=4,
\]
the class has sufficient capacity for the four two-bit messages.
The prefix encoder below selects four of these six separated words, giving the message assignments in the table. 

Set $K=n-b-1=2$, so we encode the four message indices
$m\in\{0,1,2,3\}$. At the first bit, the prefix counts are
\[
 L_H(D_{H,\beta}(0))=L_H(D_{H,\beta}(1))=3.
\]
Thus indices $0,1,2$ take the zero branch, while index $3$ takes
the one branch and becomes $0$. Continuing the prefix procedure
gives the four assignments in the table. In particular, message
$11$ is encoded as $10010100$.

To illustrate decoding, delete the first bit of $10010100$, yielding
$y=0010100$. Among the nine distinct length-eight words obtained by
inserting one bit into $y$, only $10010100$ satisfies $Hx=\beta$.
The inverse prefix procedure then recovers message $11$.

\section{Conclusion}\label{sec:conclusion}

We have given a deterministic construction of binary codes correcting any
combination of at most $t$ insertions, deletions, and substitutions in total,
with redundancy at most $2t\log_2 n+O_t(1)$ for every fixed $t\ge1$ and all
sufficiently large $n$. Construction, encoding, and decoding all run in
polynomial time in $n$ for fixed $t$, and decoding recovers the original
information message. This realizes the classical Gilbert--Varshamov guarantee
up to an additive constant without searching the exponentially large space
of binary words.

A central point is that efficient encoding does not require exact counts of
valid completions. Additive lower bounds suffice. Affine descriptions of the
errors make these bounds computable through systems of binary linear equations,
while conditional expectations provide deterministic choices of the shared
code parameters. Together, these elementary tools turn a counting argument
into an explicit construction. The approach also extends to error models with
polynomial-size, efficiently enumerable affine descriptions of confusable
pairs. Efficient decoding additionally requires efficient enumeration of the
possible transmitted words consistent with a received word.

Several questions remain. Improving the running time and its dependence on
the number of errors would help move these constructions toward practical use.
For multiple edits, it also remains to close the gap between efficient
constructions and the strongest known existence bounds. Whether the same
counting approach can reach beyond the classical Gilbert--Varshamov guarantee
is a natural direction for further work.

\section*{Acknowledgment of AI assistance}
ChatGPT-6 Astra was used during the development of this work to
assist with mathematical exploration. 
The authors take full responsibility for the correctness, originality,
and integrity of the work.

\bibliographystyle{alpha}
\bibliography{ref}

\end{document}